# Advanced interfacial phase change material: structurally confined and interfacially extended superlattice

Hyeon wook Lim[1], Young sam Kim[2], Kyu-jin Jo[3], Seok-Choi[4], Chang Woo Lee[1], Dasol Kim[1], Ki hyeon Kwon[1], Hoe don Kwon[1], Soo bin Hwang[1], Byung-Joon Choi[4], Cheol-Woong Yang[3], Eun Ji Sim[2], and Mann-Ho Cho[1]*

Keywords: interfacial phase change material, doping, vdW layer, superlattice, neuromorphic

**Interfacial Phase Change Memory (iPCM) retrench unnecessary power consumption due to wasted heat generated during phase change by reducing unnecessary entropic loss. In this study, an advanced iPCM (GeTe/Ti-$Sb_2Te_3$ Superlattice) is synthesized by doping Ti into $Sb_2Te_3$. Structural analysis and density functional theory (DFT) calculations confirm that bonding distortion and structurally well-confined layers contribute to improve phase change properties in iPCM. Ti-$Sb_2Te_3$ acts as an effective thermal barrier to localize the generated heat inside active region, which leads to reduction of switching energy. Since Ge-Te bonds adjacent to short and strong Ti-Te bonds are more elongated than the bonds near Sb-Te, it is easier for Ge atoms to break the bond with Te due to strengthened Peierls distortions ($R_{long}/R_{short}$) during phase change process. Properties of advanced iPCM (cycling endurance, write speed/energy) exceed previous records. Moreover, well-confined multi-level states are obtained with advanced iPCM, showing potential as a neuromorphic memory. Our work paves the way for designing superlattice based PCM by controlling confinement layers.**

Phase change memory (PCM) is one of the promising candidates for next-generation non-volatile memory (NVM). There have been many attempts to design superior PCMs based on single phase materials mainly composed of Ge, Sb, Te alloys[1]. However, there remains a fatal problem: the unnecessary diffusion of heat through the liquid phase generated during the RESET process due to the melt-quenching process[2]. To solve these problems, many research has been conducted on various methods such as pulse design optimization, device structure improvement, and material design to present new paradigms for designing advanced PCMs[3,4,5]. Meanwhile, outstanding device characteristics for interfacial phase change memory (iPCM) composed of alternately stacked GeTe (GT) and Sb$_2$Te$_3$ (ST) within nm ranges has been reported[6,7,8]. The main characteristic of iPCM is that two reversible phases are correspond to crystal and metastable crystal phases, not the crystalline and amorphous phases. Therefore, iPCM saves energy because it does not undergo a melting process which consumes a lot of the energy generated from heat diffusion in the liquid state during the RESET process. However, due to difficult manufacturing process and complex structure, understanding the phase change mechanism of iPCM has not sufficiently progressed. Although the phase change mechanism of iPCM has not been fully comprehended, many reports suggest that difference between the two phases is determined by local movement of atoms adjacent to interface near the vdW layer[9,10]. Some models for iPCM were suggested based on four representative structures: Petrov, inverted Petrov, Kooi, and Ferro[11]. Structures corresponding to low/high resistance states (LRS/HRS) are considered as Ferro and inverted Petrov, according to Junji Tominaga et al[12]. Despite the deficient evidence on the properties of each phase in iPCM for the switching process, advanced results which include ultralow-switching current and multi-level characteristics with the GT/ST superlattice on flexible substrates were reported[13]. However, this remarkable performance is limited to specific device structures such as pore-type structures and thermally insulating flexible substrates. Sufficient attempts have not been made to develop advanced iPCM using modified structures by doping with transition metals, such as Sc and Ti, which are already used in alloy-based PCM[14,15,16,17]. To implement PCM device using superlattice structure, further research is needed to maximize the advantages of the superlattice structure and reduce difficulties in the fabrication process.

In this study, advanced iPCM composed of alternately stacked GT and Ti-doped Sb$_2$Te$_3$ (Ti-ST) were synthesized with MBE and evaluated by various measurements. Furthermore, outstanding figure of merits were obtained with the advanced iPCM, including ultra-low power consumption (~2 pJ), high cycling endurance (~3 × 10$^9$), fast speed (~8 ns) and multi-level characteristics (~6 level). Considering structural analysis and various experimental measurements, appropriate unit cells are constructed, and density functional theory (DFT) calculations are used to investigate atomic switching characteristics within the microstructure. The structural confinement and interfacial extension of iPCM with Ti atoms enable the above advancement. Phase change process is facilitated by strengthened thermal confinement of Ti-ST blocks and the improved structural coherence near the interface. Peierls distortions of the Ge-Te bonds vicinity to Ti atoms are strengthened due to robust Ti-Te bonding and expansion of the interfacial distance by intra-structural localization of Ti-ST blocks, which also enables triggered Ge atoms to change bonding motifs through smooth atomic movement. Due to the well-controlled interfacial thermal confinement and intra-structural localization of Ti-ST layers, multi-levels with electrical pulsing are obtained. Above properties will pave the way for designing iPCM as a neuromorphic memory, in addition to their application in PCM[18,19].

Furthermore, in the case of alloy-based PCM such as GST, phase change proceeds from random positions during the phase change process. Therefore, as a method of improving phase change characteristics of these thin films, material design is used to directly modify the entire region with several dopants[14,15,16]. In this study, material design of superlattice system is optimized by modifying confinement layer (ST) to improve phase change properties in GT layer, which is the activated region surrounded by the confinement layer. Therefore, a new methodology for improving phase change characteristics of PCMs is established by controlling the confinement layer, not the activated layer. Strong features of iPCM are further improved by reducing unnecessary entropic loss during phase change process, while confinement layers of the system maintain their crystallinity. It is a new paradigm that it leads to an advancement of phase change characteristics of the entire system by improving the properties of the confinement layer surrounding changeable materials, rather than directly modifying activated region which has been conducted in alloy based PCM.

**Electrical characteristics of GT/Ti-ST superlattice device**

The structures of GST-based superlattice device and characteristic changes caused by Ti incorporation are expressed as a schematic using atomic modeling. (Fig. 1a). The feature of interfacial expansion along the vertical direction due to Ti doped in ST blocks and locally stretched Ge-Te bonds with distortion are presented. Adopting increasing step pulse programing (ISPP) method (Fig. 1b and Supplementary Fig. 2b), phase change characteristics are evaluated through device operation on the same device platform. For GT/Ti-ST SL, all figure of merits obtained by electrical evaluations are improved compared to GT/ST SL and even other PCMs (Fig. 1c-g).

Advanced iPCM's characteristics showed much better performance in cycling endurance test with a significantly higher value ($\sim 3 \times 10^9$) than the other representative phase change materials including iPCM (Supplementary Fig. 2a, c-d). Compared with GT/ST SL, it is confirmed that reversible resistance states obtained with different electrical pulses are much more confined in GT/Ti-ST SL, which is an important property that prevents each state from overlapping when various resistance states are implemented for multi-level systems. These advancements are detailed in Table 1. As Ti concentration increases above certain level, Ti-ST blocks decompose to Ti-ST and TiTe$_2$ phases occurs, which degrade the phase change characteristics (Supplementary Fig. 2e-f, Supplementary Fig3. f-g).[20] Multi-level states are also obtained by repeatedly applying different electrical pulses. It is advantageous for resistance-based memories to implement a multi-level cell (MLC) that can be used as a neuromorphic memory.[21,22] Although MLC characteristics have been implemented and reported by previous studies in the use of PCM as a resistance-based memory,[23] a gradual change in resistance caused by phase change is very difficult. Herein, well-confined multi-level resistance states in GT/Ti-ST SL are successfully implemented through cycling endurance test. Although a similar attempt was conducted with Y doped ST, only three different resistance states with unstable operating process were reported.[24]

**Enhanced properties owing to bonding characteristics with Ti**

It is observed structural modification at the atomic level after device operation through transmission electron microscopy (TEM) measurement (Fig. 2 and Supplementary Fig. 3). Fig. 2a shows the device structure with HR-TEM after applying electrical signals alternatively about 500,000 times. Fig. 2b-e shows locally amorphized region incorporated in the layered structure. It is confirmed that the parts maintaining the layered structure are strongly aligned along the c-axis direction. Septuple and nonuple Te-terminated atomic layers are observed using intensity profiling in high-angle annular dark-field scanning TEM (HAADF-STEM), which are presented in Fig. 2f-g. The hexagonal closed packed (HCP) quintuple layer (QL) structure is maintained in Ti-ST up to certain amounts of Ti incorporation. Te-terminated layers with cation intermixing can correspond to Ferro-like or Kooi-like structures[9,25]. Among the four representative models suggested by several studies on iPCM (inverted Petrov, Petrov, Ferro, and Kooi models), switching characteristics can be explained by Ferro and inverted Petrov models[9]. The internal structures of the superlattice devices after about 500,000 cycling operations show the structural change of the superlattice device. The thick Te-terminated layers inside iPCM correspond either to Kooi or Ferro structure.[26] However, Kooi structure is so stable that phase change is not possible in it.[27] Considering intensity profiled images of the asymmetric cation sites, the layers are correspond to Ferro and Ferro-like structure.

To investigate the change in the local structure of Ge atoms at the vdW interface using DFT calculations, unit cells of GT/(Ti)-ST SL were constructed using Ferro structure taking Ge/Sb intermixing[17] into account to reflect the experimental environment. Fig. 3a present unit cell structures with charge density differences in GT/(Ti-)ST, where Ti atoms substitute Sb atoms maintaining the same octahedral motif in ST[14]. Since Ge-Te bond sharing Te with strong Ti-Te bond is relatively weaker than Ge-Te bond near the Sb-Te bond, it is easier for Ge to break the Ge-Te bond near the Ti-Te and convert into the other bonding motif, which is resulted with the electrical switching properties[29]. Phase change was implemented through the change in the bonding structure of Ge atoms near cation-intermixed zone. Specifically, three relatively long bonds among the six Ge-Te metavalent bonds[30] combining in an octahedral motif are broken and then four short and strong covalent bonds are formed with tetrahedral bonds. These changes successfully explain the reversible phase change process between the metallic and insulating phases as shown in the band structures (Supplementary Fig. 4d-e). Using the amount of the occupied electron charge in the bond between Ge atoms and adjacent Te atoms, the change in structural motifs (octahedral/tetrahedral) and bond types (metavalent/covalent) before and after the phase change process can be successfully explained[31,32]. It is interesting that all elements are combined with metavalent[30,33] bonds in LRS of GT/ST SL and valence electrons are occupied with p orbital, maintaining the octahedral bonding motif. It is confirmed that p orbital alignment is well established with LRS in the GT/ST SL (Fig. 2f-g) and left side of Fig. 3c. This alignment was due to all constituent elements maintaining an octahedral motif and forming metavalent bonds with valence electrons occupied in p orbital[32]. Structural change occurs from octahedral to tetrahedral motif for Ge atoms when phase change proceeds, three relatively weak bonds among the six metavalent Ge-Te bonds in LRS are broken and new identical covalent bonds are formed through sp3 hybridization.[34] On the other hand, unlike Ge atoms, Sb atoms maintain the bonding states with octahedral motif[35]; therefore, a structural change does not occur in ST. When ST is doped with Ti, Ti maintains octahedral motif through robust Ti-Te bond substituting Sb and HCP QL structure is also well maintained. However, the original p orbital alignment is quite distorted due to spatial expansion vicinity to the interfaces as well as enhanced structural localization due to robust Ti-Te bond (Fig. 3b).

Ti atoms substituting Sb atoms pull nearest Te atoms strongly (Fig. 3), which produce relatively elongated Ge-Te bonds in the intermixed zone. Bond properties were investigated using density derived electrostatic and chemical (DDEC) 6 method (Fig. 4)[31]. Changes in bonding characteristics due to Ti doping are presented in Fig. 4a. Ti substituting Sb in the octahedral motif forms six strong and short bonds with Te, which extends the near chemical bond length or vdW gap distance (Fig. 3c-d). The localization in Ti-ST block by Ti-Te bond is strengthened (Fig. 4b, c). Thus, the more Ti atoms are involved in the cation site in the HCP

ST block, the stronger the localization within the block, resulting in increased adjacent vdW gap distance. Volume of unit cell is reduced due to the short and strong Ti-Te bond, while length along the c-axis is rather extended. For $V_2VI_3$ or $Pn_2Ch_3$ (Pn = Pnictogen, Ch = Chalcogen) materials, such as $Sb_2Te_3$ and $Bi_2Te_3$, constituting HCP QL structure, the vdW gap is significantly reduced compared with theoretically expected value due to metavalent bonding of the electrons at the interface.[36] However, shrinkage of interfacial space is relieved as interfacial metavalent bonding is decreased to some extent due to incorporated Ti in GT/Ti-ST SL. Since the phase change in iPCM occurs with limited atomic movement of Ge atoms vicinity of the interface[37], expanded interfacial space facilitate the vertical movement of Ge atom, which occurs with phase change process. These characteristics of phase change process agree well with changes in the specific bonding properties (Fig. 4). The length of chemical bonds between Ge atom located in Ge/Sb intermixed zone and Te which is pulled toward Ti is increased and the amount of electrical charges involved in the bond is decreased. (Fig. 4d, e). Ge-Te bonds near Ti atoms are easier to break by the strengthened Peierls distortion ($R_{long}/R_{short}$)[38] due to elongated Ge-Te bonds close to Ti atoms. Therefore, Ge atoms located in Ge/Sb intermixed zone are much easier for switching to the other structural motif compared with environment without Ti. Since Ge atoms bonded with adjacent Te atoms have relatively increased vacant space along the c-axis, they can be easily flipped to tetrahedral motifs. It is well understood that Ge in the center forms covalent bonds with four Te atoms with almost same bond length. The closer the bond order value is to 1, the greater the amount of the occupied electrons in the bond, which denotes the strong covalent bonding motif.[39] This tendency is more evident in GT/Ti-ST compared with GT/ST (Fig. 4f-h). Analyzing angle distribution function (ADF), it is confirmed that bond angle of Ge atoms was much more distributed around 100 degrees in GT/Ti-ST SL than in GT/ST SL, corresponding to phase change with the local structural changes of Ge atoms. Ti and Sb always maintain octahedral motifs, bond angle near 90 degrees[15,40], regardless of the structural change, while Ge atoms partially change to tetrahedral motif (covalent bonding with Te atoms) in HRS (Supplementary Fig.5). Sb always maintains an octahedral motif in GT/(Ti-)ST SL, which forms partially defective octahedral motifs regardless of phase change process, including elongated Sb-Te bonds due to Ti.[15] Ti are maintain strong octahedral motif.

**The effects on structural coherency in GT/Ti-ST superlattice**

Significant improvements in GT/Ti-ST superlattice films with Ti incorporation are also elucidated through several structural analyses. Using time-domain thermo-reflectance (TDTR) measurement, the decrease in thermal conductivity of iPCM was confirmed. It is very interesting that the reduction in thermal conductivity of 5%-doped GT/Ti-ST is larger than the 10% doped Ti-ST SL (Fig. 5a-b). This is caused by generation of $TiTe_2$ from Ti-ST block with excess amount of Ti (Supplementary Fig.3f-g). Tri-layered $TiTe_2$ blocks have large lattice mismatch about 11% with GT and ST, which deteriorate the formation of uniform interfaces in superlattice structure.[41] For GST based PCM, thermal conductivity along the cross-plane is lowered in superlattice structure than in alloy due to increased anisotropy and presence of the vdW interface.[42] Ti-ST has lower thermal conductivity than ST system. Therefore, it has more efficient intrinsic thermal barriers inside the superlattice structure when Ti is incorporated with certain amounts in ST blocks. Since PCM operates with Joule-heating process, lowered thermal conductivity of GT/Ti-ST SL helps confine the generated heat with applied electrical pulses inside the active region, which enable ultra-low energy switching for GT/Ti-ST SL.[43,44] The scheme and TEM image of superlattice device presented in Fig. 2 and the red box (Fig. 1a) illustrate the strengthened thermal confinement effect of ST blocks due to Ti incorporation with lowered thermal conductivity.

Another important role of Ti is in synthesizing an ideal iPCM by inducing the coherent growth of HCP QL ST and GT blocks by changing the process of forming HCP QL ST blocks from growth-dominant to nucleation-dominant method, which prevent deterioration of the roughness due to abrupt change of grain size during the synthesizing process[20].[20] By incorporating Ti atoms in iPCM, it is possible to synthesize an iPCM with uniform vdW layers over large areas.[20] From X-ray diffraction (XRD), X-ray reflectance (XRR), and atomic force microscope (AFM) results, it is confirmed that morphology of iPCM is significantly improved by an appropriate amount of Ti and that good crystallinity is maintained (Fig. 5c-e).

In addition, the chemical states of composed elements are studied using x-ray photoelectron spectroscopy (XPS), with an in-situ transfer system to prevent oxidation, to obtain accurate bonding state of the constituent elements, measured chemical states showed that there were no phase segregations despite Ti is incorporated and it is observed that all elements except for Te atom itself are combined with Te (Fig. 5f-i). The binding energy shift for constituents almost did not occur before and after doping. That is, the chemical circumstance of Ti bonded with Te is hardly modulated, resulting in the maintenance of the octahedral motif.[15] From above, GT/Ti-ST SL can avoid device failure due to formation of segregated phases with additional element. The atom that plays a key role in the phase change process is Ge atom (Supplementary Fig.4a-b). Therefore, the phase change mechanism in GT/Ti-ST SL is up to the behavior of Ge atoms as the same manner in GT/ST SL. Moreover, analyzing composition ratio using XPS measurement showed that the GT/(Ti-)ST SL structure has Ge-deficient GT layers (Supplementary table 1). In detail, the result indicates that intrinsic vacancies from the cation site in GT layer help Ge to move efficiently during phase change process, which is consistent with already reported research that Ge-deficient GT layers contribute to superior phase change properties in iPCM[6,45].

## Conclusions

In summary, an ultra-efficient and structurally coherent advanced iPCM was developed in this study, which led to the following figure of merit: high cycling endurance (~3 × 10$^9$), ultra-low RESET energy (~2 pJ), and high speed (~8 ns). After sufficient electrical pulsing, it is observed that internal structure of superlattice device was well maintained, i.e., Te terminated layers were preserved. The layers corresponding to Ferro structure (-Te-Ge-Te-Ge-Te-Sb-Te-Sb-Te-) contain some extent of Ge/Sb intermixing in cation sites. Interestingly, it is also confirmed that strong p orbital alignment in GT/ST SL is slightly distorted with Ti due to the strong localization. Based on Ferro structure considering Ge/Sb intermixing, bonding properties of two reversible phases in GT/(Ti)-ST SL are suggested with DFT calculation and XPS. Ge-Te bonds vicinity to Ti-Te bonds are weakened, which facilitate Ge atoms to break the bond with Te by strengthened Peierls distortion ($R_{long}/R_{short}$). Although volume size of unit cell is reduced as intra-structural localization of ST block is strengthened by incorporating Ti atoms in ST blocks, interfacial space is expanded. The phase change process is strengthened with the stable structure formed by strong intra-structural localization of Ti-$Sb_2Te_3$ blocks and switching energy consumption is reduced with enhanced thermal barrier of the blocks by lowered thermal conductivity with Ti. Consequently, Ti incorporation facilitates efficient atomic movement of Ge atoms which leads to GT/Ti-ST SL being a superior PCM.

By modifying properties of confinement layer through doping with other transition metals which enable to strengthen structural coherence and confinement effect of $Sb_2Te_3$ block of iPCM, other advanced iPCMs can be developed. Due to GT/Ti-ST SL's strong figure of merits for cycling endurance and low-energy transition by effectively reducing entropic loss during phase change process, elaborated resistance changes with repetitive electrical pulsing become possible in addition to the improvement of above phase change characteristics in single device. Advanced iPCM is expected to pave the way for developing advanced memory devices by suppressing wasted heat generated during the phase change process and effectively forming structurally coherent interfaces. Advanced iPCM has an enormous potential to be one of the strongest candidates for non-volatile memories in highly integrated memory systems such as 3D X Point structure, which is suitable to perform deep learning operations with matrix-vector-multiplication in neuromorphic computing[23,46,47].

**Table 1. Device characteristics with different PCMs**

| Atomic ratio (%) | GST alloy | Ti-ST alloy | GT/ST Superlattice | TiTe$_2$/Sb$_2$Te$_3$ (PCH) | GT/Ti-ST Superlattice |
|---|---|---|---|---|---|
| RESET Energy (pJ) | 900 | 750 | 10 | 200 | 2 |
| SET Speed (ns) | 30 | 8 | 30 | 8 | 8 |
| Cycling Endurance | $10^{5\sim8}$ | $10^{5\sim6}$ | $10^{7\sim9}$ | $2\times10^9$ | $3\times10^9$ |
| Multi-Level States (MLC) | 3 | 2 | 2~4 | 8 | 6 |

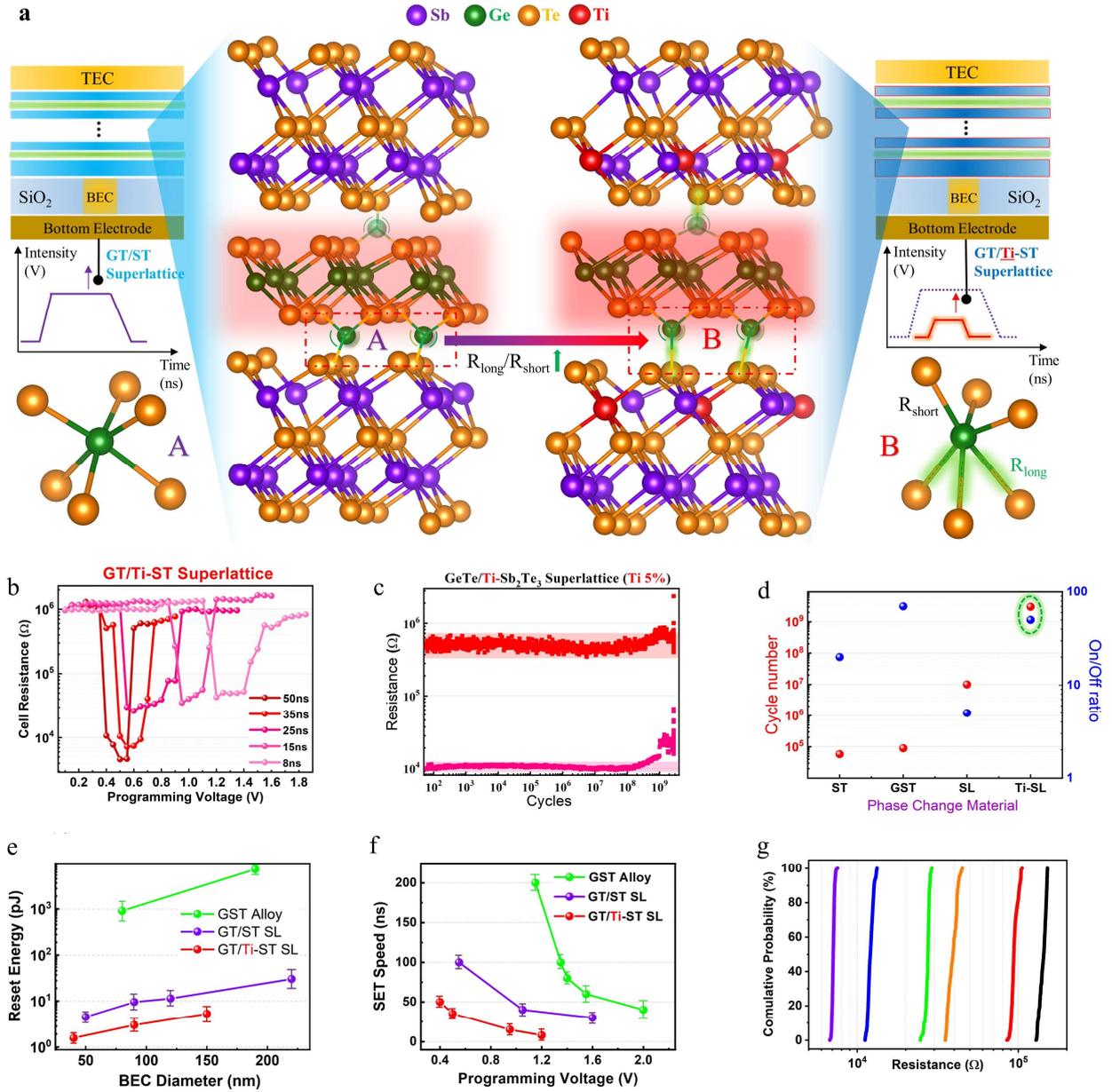

**Fig. 1. Scheme and device operation (Scheme of the superlattice structure, device and electrical data obtained from square pulses) a,** Schematics of the GT/ST SL and GT/Ti-ST SL and applied electrical pulses for the device operation are presented. The light green block represents the GeTe layer, the sky blue represents the $Sb_2Te_3$ layer, and the blue represents the $Ti-Sb_2Te_3$ layer. The following features are expressed in the central part of the scheme: The interfacial space is quite extended along the vertical direction due to Ti-ST block and the heat confinement effect of ST blocks is further strengthened with Ti incorporation, which facilitates efficient phase change process by localizing heat generated by applied electrical pulses into the GT blocks. Moreover, the diagram presents the distortion of Ge-Te bonds is strengthened as the Ge-Te bonds vicinity to Ti atoms are elongated due to short and strong Ti-Te bonds. **b,** Electrical data obtained by increasing the step pulse programming (ISPP) measurements and **c,** cycling endurance test of GT/Ti-ST SL (Ti 5%). **d,** Schematic which presents the properties obtained with cycling endurance test based on different PCMs fabricated and evaluated with same electronics. **e,** RESET energy as a function of bottom electrode contact (BEC) diameter. The above values are obtained using product of the power ($P = V^2/R$, where the cell's resistance varies) and the pulse length (t ~ ns) applied during the operation. **f,** SET speed as a function of the applied voltage bias for GST alloy and superlattice device. In obtaining above measured values of e and f. The standard deviations due to the repeated measurements and leading/trail edges of the electrical pulses are indicated by the hourglass-shaped icons. **g,** Multi-level states are obtained through the cyclability test with proper pulse design, and the population for each resistance was calculated and expressed as a cumulative probability.

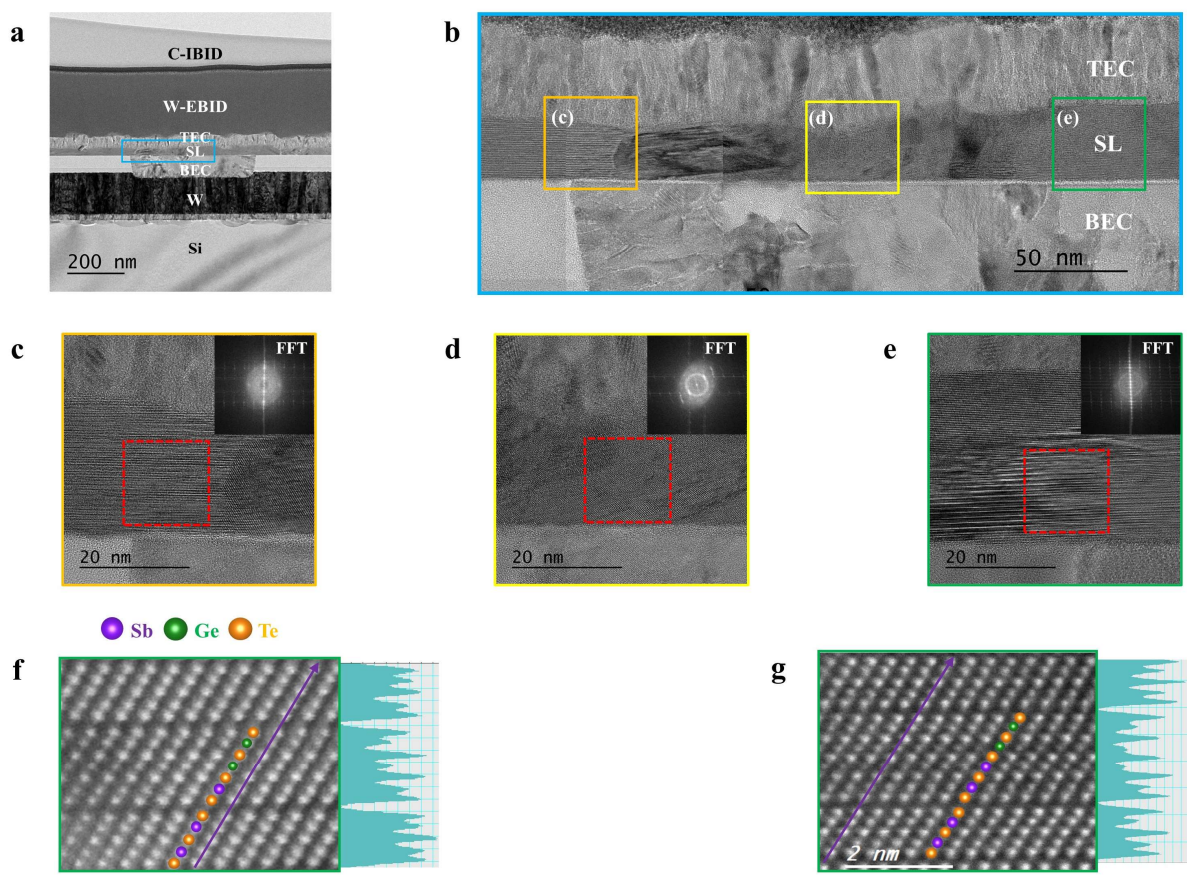

**Fig. 2. Transmission Electron Microscope (TEM) results of the superlattice device after the cyclability test. a,** TEM image of the superlattice device after cycling operations (~5 × 10$^5$ times) and **b,** magnified image of a. C-IBID denotes carbon – ion beam induced deposition and W-EBID denotes tungsten – electron beam induced deposition. **c-e,** Magnified images of the area marked with square boxes in b. The degree of crystallization and alignment of each region is presented with fast fourier transformed (FFT) image. **d,** corresponds to the amorphized region. c and e correspond to the regions where the layered structures still exist. **f-g,** is observed through HAADF-STEM by magnifying e with intensity profiling where intensity is proportional to Z$^2$. Valence electrons occupied with p orbital of constituent elements lead to atomic alignment through metavalent bonds. Te is distributed on the anion site and the other elements and vacancy are intermixed at the cation sites.

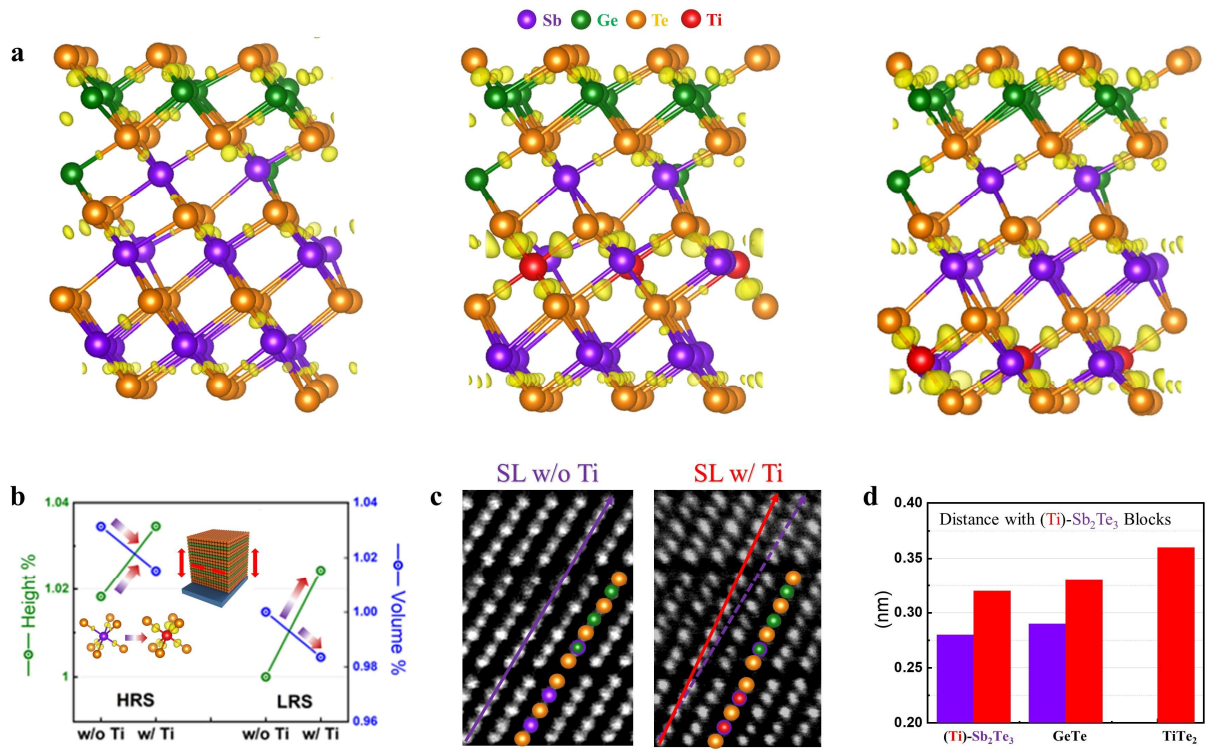

**Fig. 3. Unit cell of GT/(Ti-)ST SL and structural changes due to Ti incorporation. a,** DFT modeling of GT/(Ti-)ST SL based on Ferro structure with Ge/Sb intermixing. Strong localization due to Ti incorporation is confirmed with isosurface value being set as 0.008 e/bohr$^3$. **b,** The change in volume and length of the unit cell of GT/(Ti-)ST SL corresponded to LRS/HRS. In both reversible phases, Ti leads to reduction of unit cell volume and extension along the c-axis. **c,** HAADF-STEM images of GT/(Ti-)ST SL in LRS. Original atomic alignment is distorted with Ti incorporation. **d,** The distance (vdW gap) between (Ti-)ST block and adjacent blocks through HAADF-STEM measurement in GT/(Ti-)ST SL, purple for distances with ST and red for Ti-ST.

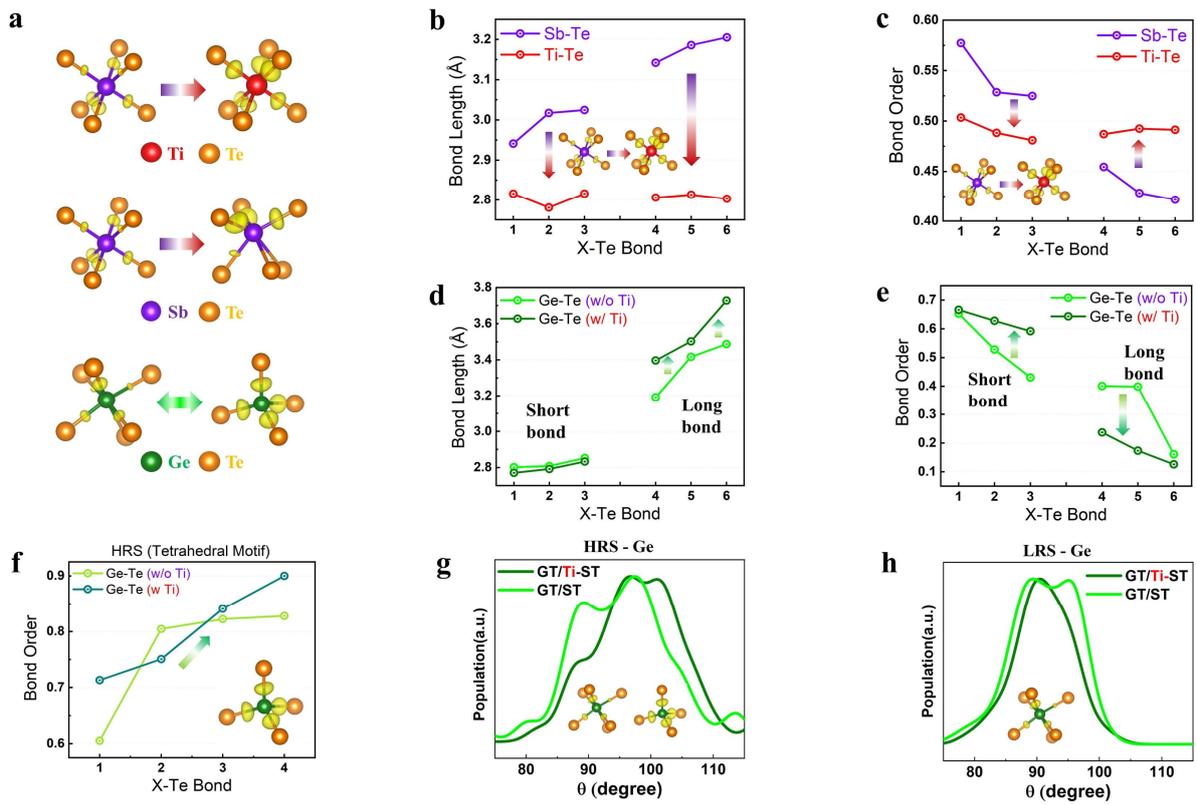

**Fig. 4. Bond properties in GT/(Ti)-ST SL. a,** Characterization of the chemical bond with the nearest Te atoms, atoms in the center remain Sb, Ti, Ge. Electron density of each atom is 0.005 e/bohr$^3$. The bond order in these systems corresponds to the value which is dividing the number of valence electrons occupied in the chemical bond between atoms by the number of atoms participating in the bond. **b,** Bond length and **c,** Bond order values for the chemical bonds where Sb and Ti at the center are bounded to Te maintaing octahedral motifs in GT/(Ti)-ST SL. **d,** Bond length and **e,** Bond order values for chemical bonds where Ge at the center is bounded to Te maintaining octahedral motifs in GT/(Ti)-ST SL. **f,** Bond order of Ge at the center are bound to Te with covalent bonding maintain tetrahedral motifs in GT/(Ti)-ST SL. **g-h,** Angle distribution function (ADF) of Ge atoms in GT/(Ti-)ST SL for two reversible phases.

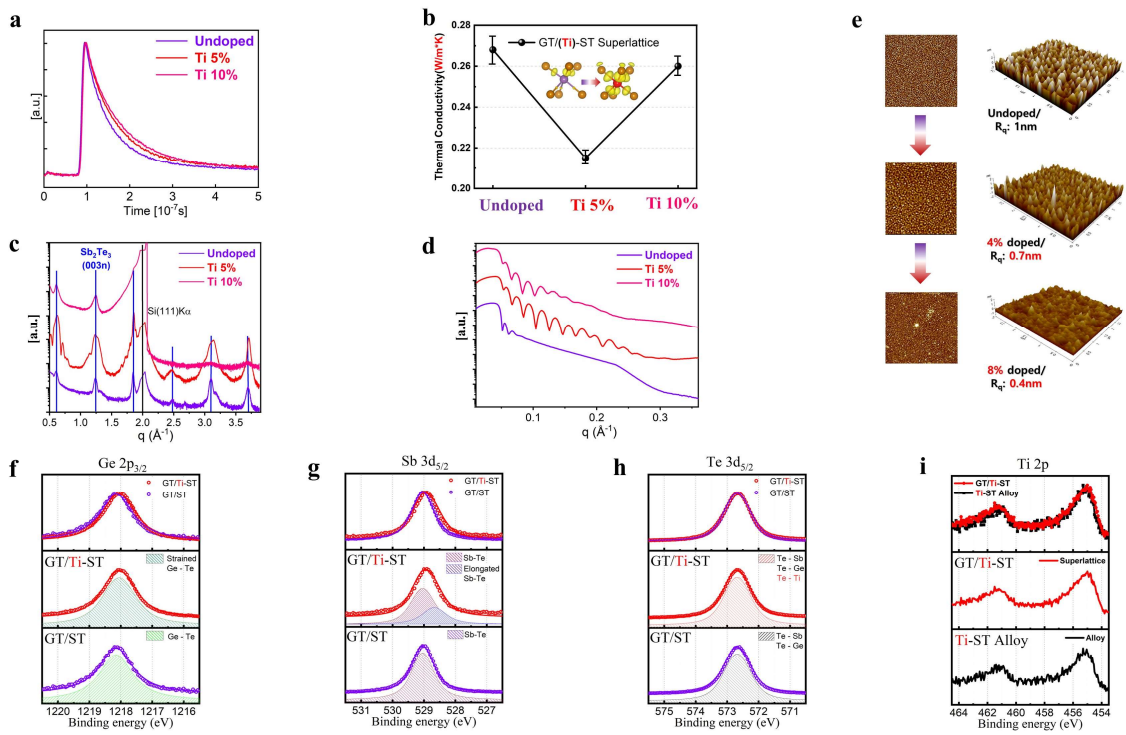

**Fig. 5. Structural analysis in GT/(Ti-)ST SL. a,** Raw data of the time-domain thermo-reflectance (TDTR) data and **b,** thermal conductivity of GT/(Ti)-ST SL. Above standard deviation occurs during the fitting operation, which is to minimize the root-mean-square error through the iterative optimization process. **c,** X-ray diffraction (XRD) and **d,** X-ray reflection (XRR) measurement of GT/(Ti)-ST SL on Si (111) substrate through θ-2θ method. **e,** Atomic force microscope (AFM) data of GT/(Ti)-ST SL on Si (111) substrate. X-ray photoelectron spectroscopy (XPS) data of each element, which corresponds to **f,** Ge, **g,** Sb, **h,** Te, **i,** Ti of GT/(Ti)-ST SL and Ti-ST alloy. Sb 3d peak is broadened asymmetrically because Sb $3d_{5/2}$ peak maintaining octahedral motif is deconvoluted to two peaks corresponded to elongated and original bond due to short and strong Ti-Te bonds, which is consistent with the results of the micro-structural change (Fig. 4d-e).

**Methods**

**Sample Preparation.** The thin film preparation process for structural analysis is as follows: iPCM is synthesized with MBE. In the case of molecular beam epitaxy, the substrate is fixed at the center of the upper range under high vacuum (base pressure ~ high $10^{-9}$ Torr) and the constituent elements of the material to be deposited are evaporated in gaseous form onto the substrate through thermal evaporation. When synthesizing the film on the silicon substrate, which is oriented in the (111) direction, the substrate (silicon wafer) is soaked in acetone and isopropanol successively for a few minutes to desorb organic matter, ultrasonicated, and soaked in a buffered oxide etchant to remove the native oxide ($SiO_2$) of the substrate, and finally, rinsed with deionized (DI) water. Then, the Si wafer is transferred to the MBE chamber and annealed at a high temperature (about 600 °C) to remove the remaining impurities, cooled to room temperature, and then Ti-ST is deposited at room temperature with a thickness of 3 nm. Heat treatment at 190 °C (for the 5% doping; the higher the doping percentage, the higher the required heat treatment temperature) for 30 minutes follows, and then 4~5 nm of Ti-ST is stacked while maintaining the above temperature. A robust seed layer is formed through the above two-step process. After lowering the temperature slowly to about 20 °C, the temperature of the effusion cell containing each element is adjusted to evaporate each source with the intended ratio (1.0 Å/min for Ge, 1.5 Å/min for Sb, 3 Å/min for Te, and about 0.4 Å/min for Ti). Each source (Ge, Sb, Te) is heated in an isolated Knudsen cell containing a cylindrical crucible, and then a beam of each component directed toward the substrate is ejected from the cell and adsorbed on the surface. Ti is evaporated with an e-beam evaporator where a hot electron beam is generated from the filament and accelerated by an electric/magnetic field; it hits the source rod to vaporize Ti from the source and then transfer it to the substrate. Each material has a different desorption rate depending on the selectivity of the substrate and the pre-deposited material. Considering this, the deposition rate on the substrate surface is precisely controlled through the Proportional-Integral-Differential (PID) controller to enable the precise control that responds to every moment for the growth of a very high-quality thin film. After synthesizing the seed layer and lowering the temperature to about 20 °C, approximately 1 nm of the GT layer and approximately 2~3 nm of Ti-ST is deposited alternately. The above process is repeated until the top layer (Ti-ST) is deposited.

**X-Ray Diffraction/Reflection (XRD/XRR).** The XRD measurement is performed using a high-resolution X-ray diffractometer with a 9 kW Cu Kα (wavelength ~ 1.5406 Å) radiation source. The measurement was conducted with the Po-hang accelerator 5D beam line.

**Raman Spectroscopy.** Raman spectra were obtained through micro-Raman spectroscopy (Horiba Lab Ram ARAMIS) using a 532 nm wavelength Nd: YAG laser with a 100x objective and 2400 grooves/mm grating. The spectra were calibrated at a silicon peak of 520 cm$^{-1}$.

**X-ray photoelectron spectroscopy (XPS).** In-situ XPS spectra of the core levels were measured using the PHI 5000 Versa Probe made by ULVAC-PHI, where the base pressure was about 4 x 1$^{-10}$ Torr. The XPS spectra were obtained using monochromatic Kα radiation with an analyzer pass energy of 23.5 eV, providing an overall experimental resolution of 200 meV. The background of each spectrum was subtracted by the Shirley method. For the depth profile, a 500 kV Ne$^+$ ion beam was used, a modest condition to prevent transformation of the samples.

**Keithley (I-V test).** To probe the operation of the test cells, an Agilent 33600A pulse generator and a Keithley 2636B source-meter were used for applying electrical pulses and measuring the resistance during the switching process, respectively. To switch the cells between several reversible phases, electrical pulses which were 8~200 ns in width and 0.1~3 V in height were applied, and each state's resistance was measured at 0.1 V.

**Transmission Electron Microscope (TEM).** TEM specimens for cross-sectional observation of the super-lattice devices were prepared using a focused ion beam (FIB, NX2000; Hitachi Inc., Japan) instrument. To maintain the initial state of the specimen, tungsten and carbon were used as protective layers. During the FIB process, the TEM specimen was etched with Ga$^+$ ions at 30 kV, 10 kV, and 5 kV, successively. The TEM images were taken at an accelerating voltage of 200 kV using a JEM-ARM200F (JEOL Ltd.) and One View camera (Gatan Inc.). The STEM images were acquired at an accelerating voltage of 200 kV using an aberration-corrected STEM (JEM-ARM200F, JEOL Ltd.). STEM images were collected using a convergence semi-angle of 21 mrad and a high-angle annular dark field (HAADF) detector with a collection semi-angle of 68-280 mrad.

**Density Functional Theory (DFT) Calculation.** The electronic structures of the GT/(Ti-)ST SL were calculated using DFT with the Vienna Ab initio Simulation Package (VASP).[48] All calculations were implemented at the generalized gradient approximation level, using the Perdew-Burke-Ernzerhof (PBE)[49] functional. The Grimme's D3[50] dispersion correction term was added to capture long-range interactions during structure optimization. The Brillouin zone was sampled with 11 × 11 × 11 (21 × 21 × 5) k-points, the kinetic energy cutoff for the plane wave was set to 500 eV (400 eV), and the convergence criterion for the self-consistent loop was 1E-6 eV (1E-4 eV) for optimization (single-point calculation) until the difference in force between the steps was less than 0.01 eV/Å. The superlattices were relaxed with 2 × 2 × 2 Brillouin zone sampling and gamma point sampling, respectively. By fabricating several 3 × 3× 1 supercells of ST, representative structures were created in which Sb and vacancies were mixed in a ratio of 2:1 in all layers except for the Te layers. The exact location of the vacancy was determined by the CD and monitored during the process of site switching and optimization. Subsequently, the CD near the vacancy site was numerically integrated. The superlattices were constructed in the same manner. The vacancies, Sb, and Ge were mixed at the interface between GeTe and ST. The bond order was calculated using the density derived electrostatic and chemical (DDEC6) approach.[31] The band structure calculations were carried out in the reciprocal space along the symmetry points (K, Γ, M).

**Time-Domain Thermo-Reflectance (TDTR).** To measure the thermal metrology based on thermo-reflectance, the N8- 200 made by TMX Scientific Transometer™ was used. A pulsed heating laser of 4.7–5.8 μJ was illuminated to change the temperature of the surface and heat reflection energy of the sample with the film stack. At the same time, a probing laser (18.5–23.3 mW) was used to record the change in light reflectivity (i.e., thermal decay) as a function of time. Two unknown independent parameters can be extracted after fitting by combining the experimental data with known material parameters. These unknown parameters can be calculated using the coefficient of thermal reflectance, representing the relationship between temperature and reflectivity changes. During the fitting process, optimal calculations were performed through an iterative optimization process that minimized the root-mean-square error between the experimental and numerical decay curves. Eventually, the inverse numerical solution can acquire the film stack's thermal conductivity and boundary resistance.

**Acknowledgements**

This research was supported by grants from the Government of Korea (MSIP) (No. 2021M3H4A1A03052566) and the Korea Semiconductor Research Consortium (KSRC) through a project developing source technologies for future semiconductor devices and NRF-2020R1A2C2007468 for DFT Calculations and the Korean government (Grant No. 2017R1A5A1014862, SRC program: 22 vdWMRC (center) and supported by the National Research Foundation of Korea (NRF) grant funded by the Government of Korea (MSIP) (No. 2021M3H4A1A03052566).

**Author Information**

Authors and Affiliations

H. Lim, C. Lee, D. Kim, K. Kwon, H. Kwon, S. Hwang, M. Cho

Department of Physics, Yonsei University, 50 Yonsei-ro, Seoul 03722, Republic of Korea

E-mail: mh.cho@yonsei.ac.kr

Y. Kim, E. Sim

Department of Chemistry, Yonsei University, 50 Yonsei-ro, Seoul, Republic of Korea

K. Cho, C. Yang

School of Advanced Materials of Science and Engineering, Sungkyunkwan University, 2066 Seobu-ro, Suwon, Gyeonggi-do 16419, Republic of Korea

S. Choi, B. Choi

Department of Materials Science and Engineering, Seoul National University of Science and Technology 232, Seoul, Republic of Korea

Supporting Information

# Advanced interfacial phase change material: structurally confined and interfacially extended superlattice

Hyeon wook Lim[1], Young Sam Kim[2], Kyu-jin Cho[3], Seok -Choi[4], Chang woo Lee[1], Da sol Kim[1], Ki hyeon Kwon[1], Hoe don Kwon[1], Byung -Joon Choi[4], Cheol-Woong Yang[3], Eun Ji Sim[2], and Mann-Ho Cho[1,2,]*

The supplementary information provides additional information about the manuscript. Supplementary explanations for the preparation process, device characterization, evaluation, and optimization process, and the structural analysis presented in the manuscript are presented here with additional data. The supplementary information is presented below to strengthen the thesis of the manuscript and help understand the system of advanced iPCM.

Firstly, the fabricating process of a single device after the thin film synthesis involves very meticulous detail. In the case of superlattice based PCM, unlike alloy based PCM, it is necessary to synthesize thin film maintaining crystallinity to keep the layered structure. Therefore, the fabricating process is not filling PCM material in a hole made with photoresist material but synthesizing thin film with sufficiently high temperature annealing in large area and then isolating it through an etching process to manufacture the device. In this process, organic material including photoresist material became hardening and some decrease in the on/off ratio in electrical operating is caused by this residual organic matter due to hardening. During the dry etching process with the plasma system, there was a thermal hardening issue with photo-resist materials. For bottom electrode contact (BEC), it affects the thin film's initial growth due to its direct contact with it, and it is a place where electrical pulses are converted into the Joule heating process due to high current density. Therefore, it is hard to change the type of BEC. For synthesizing the iPCM on the device platform, a thin seed layer of approximately 3 nm is deposited at room temperature and treated with a post annealing process as a buffer layer to minimize the difference in thermal properties and selectivity between $SiO_2$ and BEC (TiN), and then a solid seed layer is deposited with an additionally deposited seed layer through the co-dep process. In the case of the top metal, because it acts as a thermal reservoir, which is a passage where the heat generated from the BEC passes upward after the phase change of the thin film, there are relatively few restrictions, including the difference in work function with the thin film compared to the selection of BEC. Considering the above characteristics, iPCM and top metal are sequentially synthesized over the BEC structure, and the isolation process is then performed with a uniform pattern successively, which is presented in Supplementary Fig.1.

From the result of the endurance test, it was confirmed that by increasing Ti composition in iPCM, the on/off ratio and cycle limit decreased. In particular, it is confirmed that device degradation occurs rapidly with highly doped systems (>10%). This degradation is due to segregation of the $TiTe_2$ block caused by excessively incorporating Ti. The $TiTe_2$ block has a quite large lattice mismatch with GT and ST blocks of about 11%, far exceeding the low lattice mismatch between GT and ST (~2.2%). In large areas above the Si substrate, it is confirmed that the vdW material ST and the non-vdW material GT are alternately and uniformly stacked on the ST seed layers. Even though Ti element is incorporated, it is confirmed that the original crystal structure of iPCM is well maintained. However, the segregation phase of the HCP $TiTe_2$ tri-layers appears with excessive doping (>10%).

The changes in the binding energy of the core electron in Ge atoms due to the change of the bonding motif in Ge atoms during the phase change process are confirmed, which is not found in the other elements during the phase change process. When the band structure for the phases before and after the phase change is calculated through the local bonding motif changes of Ge atoms near the interface, the Fermi level crosses through the edge side of the valence band in the LRS, but it does not cross any band in the HRS, specifically, lying between the band gaps. Therefore, it is confirmed that the metal to insulator transition occurred. The Raman spectroscopy results show that some contributions of GT and GST occurred in the phonon mode from the phase change process, in addition to the phonon mode corresponding to the $Sb_2Te_3$, which continues to exist before and after the phase change process. In Raman spectroscopy, the phonon modes of ST were dominant in LRS but the corresponding phonon mode decreased and the phonon mode corresponded to GST increased in HRS compared to LRS. However, when phase separation occurs by irradiating much stronger laser than intensity suitable for RESET, stable $TiTe_2$ structures are formed and it deviates from reversible phases.

Moreover, changes in the bond angle for the constituent elements in iPCM according to the local structural change of Ge atoms are also presented in Supplementary Fig.5. Each angular region is divided into tetrahedral and (defective) octahedral sections. It is confirmed whether the bonds correspond to each bonding motif with indicators such as charge density, bond length, and bond order. In the case of Sb, the defective octahedral structure is observed in the doped system much more than in the undoped one due to Ti substituting some of the Sb atoms. In the process of changing from LRS to HRS, the degree of change in the bonding motif is larger in doped system than in the undoped one compared to the existing octahedral motif of Ge element.

**Table S1. Atomic concentration (in-situ XPS)**

| Atomic ratio (%) | Ge | Sb | Te | Ti |
|---|---|---|---|---|
| GT/ST SL | 9.5 | 30.5 | 60 | - |
| GT/Ti-ST SL | 8.5 | 27 | 59.5 | 5 |

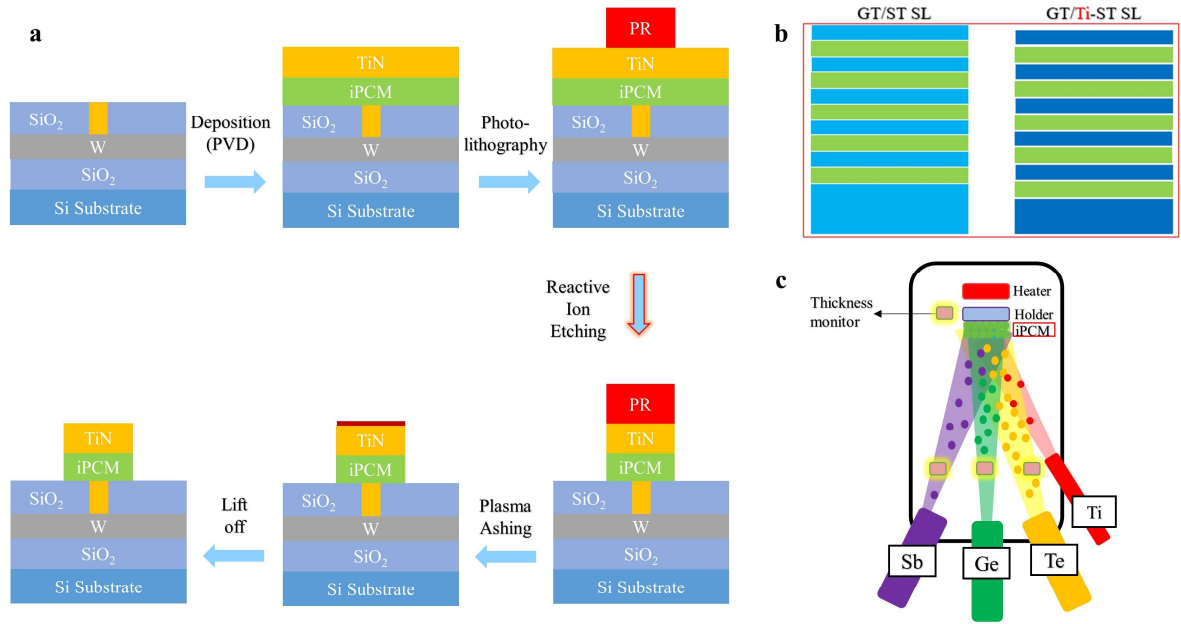

**Supplementary Figure 1: Schematic fabrication process of the superlattice device. a,** A scheme of the whole process up to the fabricating the superlattice PCM device. Since continuous annealing is conducted to maintain high crystallinity for synthesizing superlattice based PCM, the fabrication process is performed by firstly synthesizing in large area and then vertically etching the unnecessary parts, rather drilling a photoresist hole and then filling the thin film in it and progressing the lift-off. **b,** A scheme of the GT/(Ti-)ST SL on Si (111). Due to the high structural coherence of Ti-ST, robust superlattice deposition was possible in GT/Ti-ST SL even above the synthesis of thinner seed layers. **c,** A scheme of the molecular beam epitaxy (MBE), which is a deposition chamber. Each source is transported in vapor form through thermal evaporation wih the effusion cell for GST and this process was delicately controlled through the PID loop, which enable the detailed deposition with Angstrom scale. In the case of Ti, since the evaporation temperature is high, it was deposited with thermal evaporation method by hitting the rod source with hot electrons generated from the filament through high voltage/current.

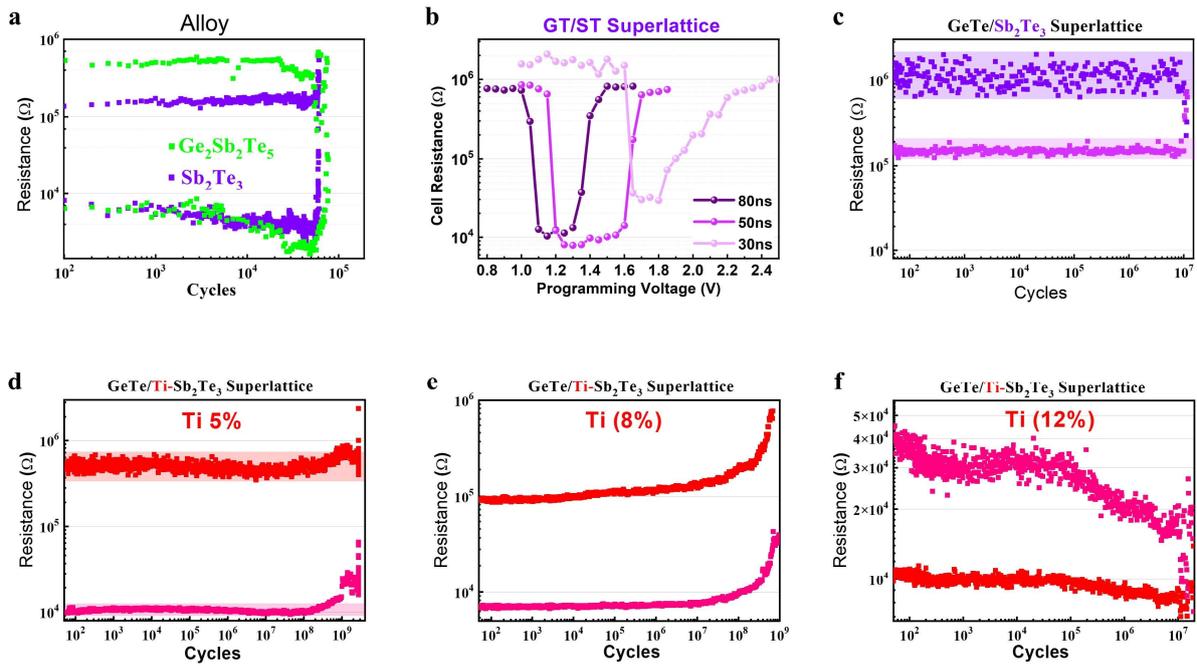

**Supplementary Figure 2: Electrical evaluation of several PCM including iPCM.** Cycling endurance of **a,** alloy-based phase change materials and **b,** electrical data of GT/ST SL obtained with the ISPP method. **c,** GT/ST SL and **d-f,** GT/Ti-ST SL varying the degree of Ti concentration in the superlattice device. The cycle limit decreased and the tendency to maintain the confined resistance levels collapsed dramatically by increasing Ti concentration ratio higher than 5% in GT/Ti-ST SL.

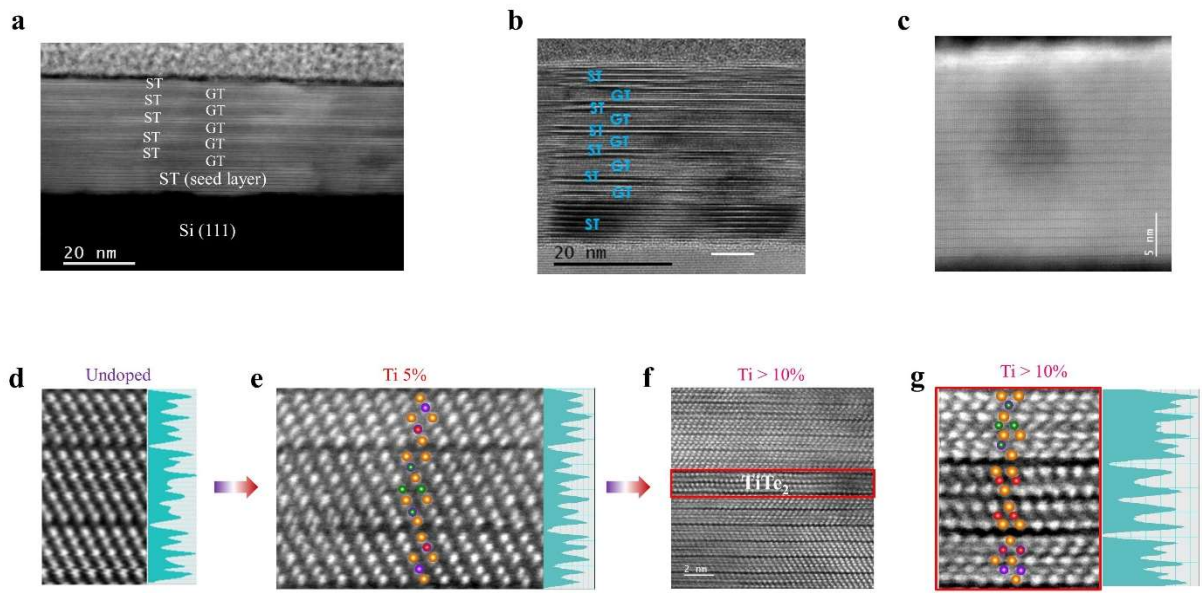

**Supplementary Figure 3: TEM image of the superlattice over a large area and HAADF-STEM image as a function of Ti (%).** **a-c,** High resolution TEM images of the iPCM. Following images are HAADF-STEM image of **d,** GT/ST SL and **e** GT/Ti-ST SL (Ti 5%) and **f,** GT/Ti-ST SL (Ti > 10%) where intensity profiling is conducted. **g,** is magnified image of f.

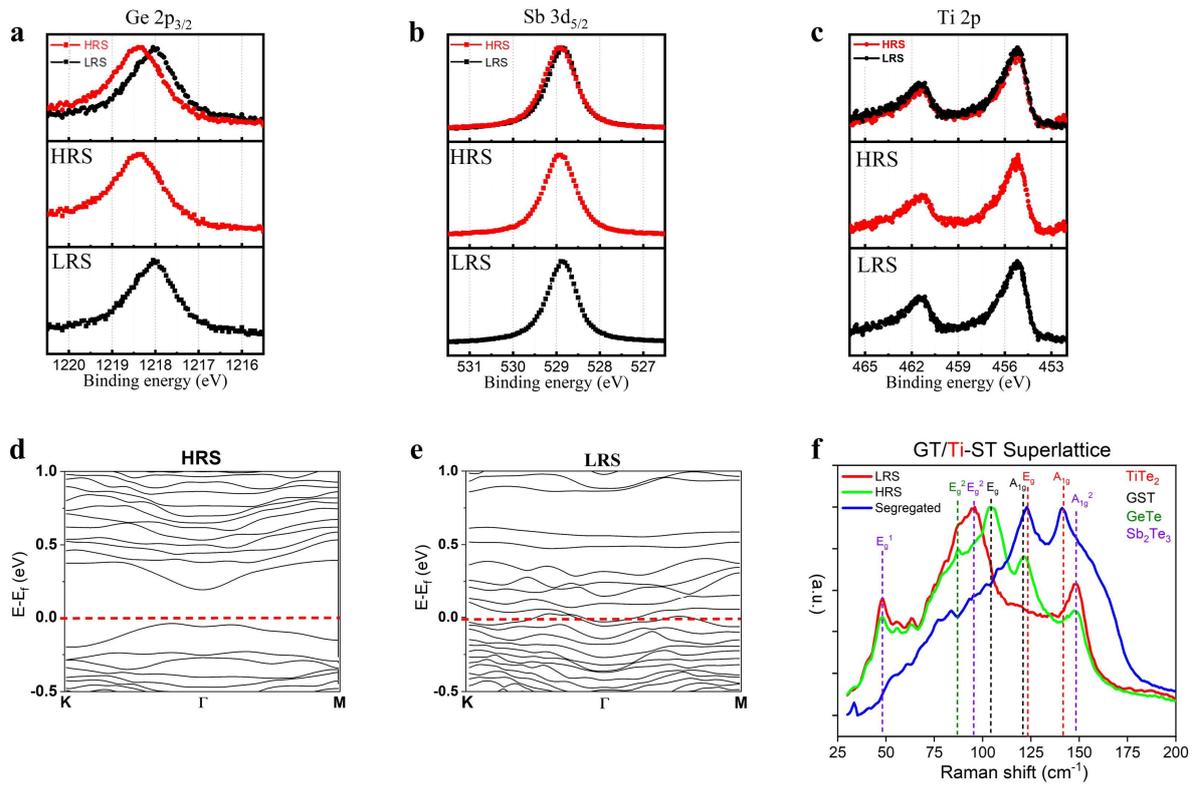

**Supplementary Figure 4: Structural analysis and band structure before and after phase change process.** Binding energy of cation atoms: **a,** Ge, **b,** Sb and **c,** Ti are obtained with XPS measurement. When the optical coefficient and electrical resistance of GT/Ti-ST SL were changed through the irradiating pulsed optical laser to films, only binding energy of Ge atom was changed. High binding shift occurred due to increase in covalent bonding from octahedral to tetrahedral transition in Ge-Te bond. From the thin film synthesis process to optical laser irradiation and XPS measurement, all processes were carried out through in-situ process.[28] A high binding shift occurred d, e Band structures of the reversible phases are presented. It was confirmed that band crossing of the fermi level did not occur in HRS and it occurred when more than a certain ratio of tetrahedral Ge atoms which have covalent bonding with Te were generated. **f,** Phonon mode of the GT/Ti-ST SL before and after phase change and segregated phases, measured with raman spectroscopy.

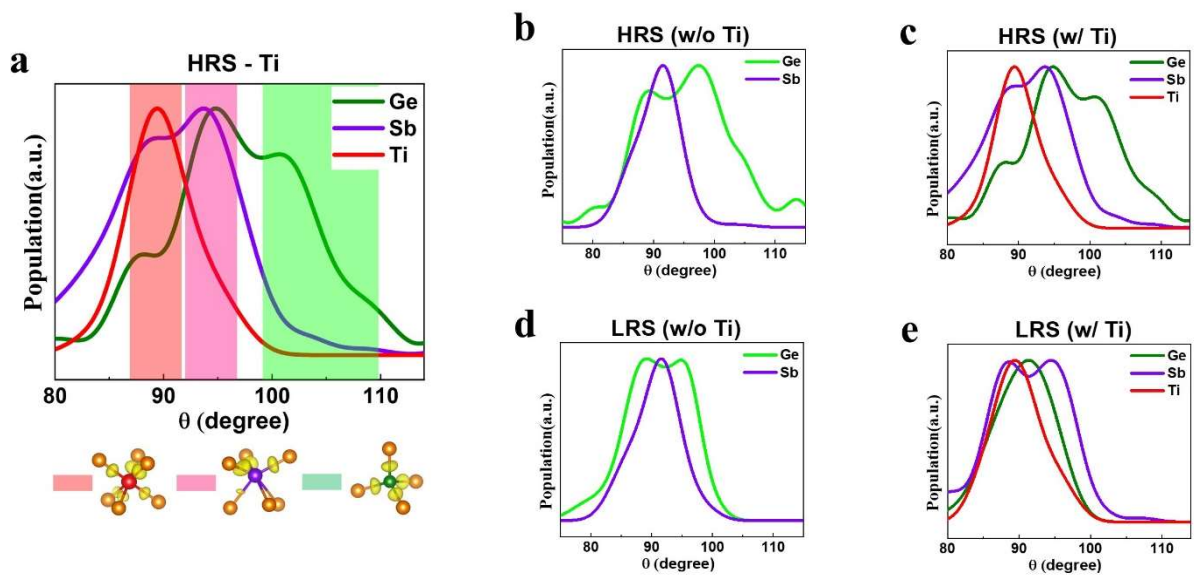

**Supplementary Figure 5: Angle distribution function (ADF) in GT/(Ti)-ST SL before and after the phase change process. a,** ADF of cation atoms in GT/Ti-SL, the colored areas indicate the chemical bonds that combined with the octahedral, defective octahedral, and tetrahedral motifs from the left side, respectively. **b-e,** ADF of the Ge, Sb, and Ti atoms in the GT/(Ti)-ST SL.